\documentclass[11pt]{article}
\usepackage{hyperref}
\pdfoutput=1

\begin{document}
\title{Fluid Mixing from Viscous Fingering}
\author{Birendra Jha \\
Luis Cueto-Felgueroso \\
Ruben Juanes
\\\vspace{6pt} Civil and Environmental Engineering
\\ Massachusetts Institute of Technology, Cambridge, MA 02139, USA}
\maketitle

\begin{abstract}
	Viscous fingering is a well-known hydrodynamic instability that sets in when a less viscous fluid displaces a more viscous fluid~\cite{homsy87-fing}. When the two fluids are miscible, viscous fingering introduces disorder in the velocity field and exerts a fundamental control on the rate at which the fluids mix. We present a fluid dynamics video of the mixing process in a viscously unstable flow, generated from a high-resolution numerical simulation using a computational strategy that is stable for arbitrary viscosity ratios. We develop a two-equation dynamic model of concentration variance and mean dissipation rate to quantify the degree of mixing in such a displacement process~\cite{jhacueto11-prl}. The model reproduces accurately the evolution of these two quantities as observed in high-resolution numerical simulations and captures the nontrivial interplay between channeling and creation of interfacial area as a result of viscous fingering.
\end{abstract}

\section{Introduction}
Mixing efficiency at low Reynolds numbers can be enhanced by exploiting hydrodynamic instabilities that induce heterogeneity and disorder in the flow. The unstable displacement of fluids with different viscosities, or viscous fingering~\cite{homsy87-fing}, provides a powerful mechanism to increase fluid-fluid interfacial area and enhance mixing. We numerically simulate the process of viscous fingering in a two-dimensional porous medium for two perfectly miscible, incompressible, neutrally buoyant fluids. The physical model describes
flow in a porous medium or Hele-Shaw cell---a thin gap between two parallel plates. The goal of the simulation is to demonstrate and understand mixing between the two fluids due to viscous fingering. The mathematical model is comprised of the advection-dispersion equation, Darcy's law, and the divergence-free velocity condition. There are two governing parameters in the non-dimensionalized set of equations--viscosity contrast between the two fluids and the P\'eclet number of the flow (ratio of advection to diffusion). The fluid dynamics video accompanying this document is created from a simulation where the viscosity ratio is $33$ and the P\'eclet number is $10000$.

We will describe the first simulation movie in the accompanying video file. This simulation is generated using snapshots of the concentration field at successive time steps. Light yellow color indicates the less viscous fluid and dark color indicates the more viscous fluid. Intermediate colors denote a mixture of the two pure fluids. We have used an exponential viscosity-concentration relationship for the mixture, i.e., viscosity of the mixture is proportional to the exponential of concentration of the more viscous fluid in the mixture. The displacement is from left to right and the reference frame is moving with the average speed of the flow. A slightly perturbed flat interface separates the two fluids initially. Periodic boundary conditions are applied in the transverse direction. Fresh less viscous fluid enters the domain continuously from the left boundary and the right edge is a natural flow boundary.

The less viscous fluid tries to move faster than the more viscous fluid because of its higher mobility. The displacement front becomes unstable leading to the formation of viscous fingers. With time, these fingers grow longitudinally much faster than they can diffuse transversely. This leads to stretching of the initial interface between the two fluids. Tips of these fingers split and smaller fingers emerge from them continuously, leading to further stretching of the interface and overall higher mixing rate across the interface. Other interesting nonlinear mechanisms visible during the fingering process are shielding of one finger by another, merging of fingers, fading of fingers and also channeling of a dominant finger through the more viscous domain. These events result from the non-local coupling between the concentration and the pressure fields due to a concentration-dependent viscosity.

We show that viscous fingering leads to two competing effects. On one hand, it enhances mixing by inducing disorder in the velocity field, and increasing the interfacial area between the fluids. On the other, it causes channeling of the low viscosity fluid, which bypasses large areas of the flow domain--these regions remain unswept thereby reducing the overall mixing efficiency. This competition between creation of fluid-fluid interfacial area and channeling results in nontrivial mixing behavior. We develop a two-equation dynamic model for evolution of the concentration variance and mean scalar dissipation rate to quantify the
degree of mixing in a viscously unstable displacement~\cite{jhacueto11-prl}. We use our analysis to predict the range of viscosity contrast that maximizes mixing.

See http://juanesgroup.mit.edu/publications/mixfing for additional videos of mixing induced by viscous fingering.

\bibliographystyle{unsrt}
\bibliography{MixingViscousFingering_FluidDynamicsVideo_Report}

\end{document}